\documentclass{PoS}

\usepackage{graphicx}
\usepackage{multirow}
\usepackage[nice]{nicefrac}
\usepackage{amssymb, amsmath}  

\newcommand{\bi}{\begin{itemize}}
\newcommand{\ei}{\end{itemize}}

\newcommand{\be}{\begin{equation}}
\newcommand{\ee}{\end{equation}}
\newcommand{\bea}{\begin{eqnarray}}
\newcommand{\eea}{\end{eqnarray}}
\newcommand{\op}[1]{\ensuremath{\langle B_q^0 | \mathcal{O}_{#1} | \bar{B}_q^0 \rangle (\mu)}}
\newcommand{\me}[1]{\ensuremath{\langle B_q^0 | \mathcal{O}_{#1} | \bar{B}_q^0 \rangle}}

\newcommand{\bt}[1]{\begin{table}[h!]\begin{tabular}{#1} \hline\hline  \\[-0.5em]}
\newcommand{\et}[2]{\hline\hline \end{tabular} \caption{#1} \label{#2} \end{table}}

\def\cpt{\raise0.4ex\hbox{$\chi$}PT}
\def\scpt{S\raise0.4ex\hbox{$\chi$}PT}
\def\rscpt{rS\raise0.4ex\hbox{$\chi$}PT}

\title{Neutral $B$ mixing from $2+1$ flavor lattice QCD}

\ShortTitle{$B_d^0$ and $B^0_s$ mixing}

\author{
E.D.~Freeland$^{\, *\ a}$
C.M.~Bouchard$^{\ b}$,
C.~Bernard$^{\ c}$,
A.X.~El-Khadra$^{\ d}$,
E.~G\'amiz$^{\ e}$,
A.S.~Kronfeld$^{\ f}$,
J.~Laiho$^{\ g}$, and
R.S.~Van~de~Water$^{\ f}$
\hphantom{\speaker{E.D.~Freeland}}
\\ \\
\llap{$^a$}Department of Physics, Benedictine University, Lisle, IL 60532, USA\\
\llap{$^b$}Department of Physics, The Ohio State University, Columbus, OH 43210, USA\\
\llap{$^c$}Department of Physics, Washington University, St. Louis, MO 63130, USA\\
\llap{$^d$}Physics Department, University of Illinois, Urbana, IL  61801, USA \\
\llap{$^e$}CAFPE and Departamento de Fisica Teorica y del Cosmos, Universidad de Granada,
E-18002 Granada, Spain\\
\llap{$^f$}Theoretical Physics Department, Fermi National Accelerator Laboratory,~Batavia, IL  60510, USA\\
\llap{$^g$}SUPA, School of Physics and Astronomy, University of Glasgow, Glasgow, G12 8QQ, UK}
\author{Fermilab Lattice and MILC Collaborations\\
Email: \email{eliz@fnal.gov}}

\abstract{We present an update of the Fermilab-MILC Collaboration's calculation of hadronic matrix elements for $B^0$-$\bar{B^0}$ mixing.  
This work is a more extended analysis than our recent publication of the SU(3)-breaking ratio~$\xi$~\cite{Bazavov:2012zs}.
We use the asqtad staggered action for light valence quarks in combination with the Fermilab interpretation of the Sheikoleslami-Wohlert action for heavy quarks.  The calculations use MILC's 2+1 flavor asqtad ensembles.  Ensembles include four lattice spacings from approximately 0.125~fm to 0.045~fm and up/down to strange quark mass ratios as low as 0.05.  Our calculation covers the complete set of five operators needed to describe $B$ mixing in the Standard Model and beyond.  In addition to an update including a fuller set of analyzed data, we comment on the form of the staggered \cpt\ extrapolation function.}

\FullConference{The 30th International Symposium on Lattice Field Theory - Lattice 2012,\\
		June 24-29, 2012\\
		Cairns, Australia}

\begin{document}

\section{Introduction}

A primary focal point in the search for new physics is the precision determination of the  Cabibbo-Kobayashi-Maskawa (CKM) matrix, 
which describes quark flavor physics.
One process of great relevance in this endeavor is neutral $B$ mixing.
In the Standard Model, this mixing occurs via a loop process where the contributions 
are suppressed both by CKM matrix elements, and, for all but the top quark, masses of the loop quark. 
Therefore, new physics could present a relatively large signal.
It has been argued that recent tension between the Standard Model and flavor physics experiments could be alleviated by the presence of new physics in $B$ mixing~\cite{Laiho:2009eu, Lunghi:2009ke, Laiho:2011nz, Lenz:2010gu, Bona:2008jn, Lenz:2012az}.
While the latest analyses indicate that this may not be the case~\cite{CKM-Nierste}, the future will, no doubt, bring new twists, and precise calculations of the theoretical inputs to $B$ mixing are necessary for a thorough understanding of quark flavor physics~\cite{CKM-Summary-HM}.

To leading order in the operator product expansion, the most general $\Delta B = 2$ effective Hamiltonian contains eight dimension-six operators
\be
	\mathcal{H}_{\rm eff} = \sum_{i=1}^5 C_i \mathcal{O}_i  + \sum_{i=1}^3 \tilde{C}_i \tilde{\mathcal{O}}_i ,  \label{eq:hamiltonian}
\ee
with
\be
	\begin{array}{l l}
	\mathcal{O}_1 = (\bar{b}^\alpha \gamma_\mu L q^\alpha )  \; (\bar{b}^\beta \gamma_\mu L q^\beta ),	&
	\mathcal{O}_4 = (\bar{b}^\alpha L q^\alpha )  \; (\bar{b}^\beta R q^\beta ), \\
	\mathcal{O}_2 = (\bar{b}^\alpha L q^\alpha )  \; (\bar{b}^\beta L q^\beta ),	&
	\mathcal{O}_5 = (\bar{b}^\alpha L q^\beta )  \; (\bar{b}^\beta R q^\alpha ),	\\
	\mathcal{O}_3 = (\bar{b}^\alpha L q^\beta )  \; (\bar{b}^\beta L q^\alpha ),
	\end{array}  \label{eq:susybasis}
\ee
where $\alpha$ and $\beta$ are color indices, $L$ and $R$ are 
the projection operators $\frac{1}{2}( 1 \pm \gamma_5)$, and $\tilde{\mathcal{O}}_{1,2,3}$ 
are obtained from $\mathcal{O}_{1,2,3}$ by $L \to R$.
%
Due to parity conservation in QCD, the matrix elements 
$\langle B_q^0|\tilde{\mathcal O}_{1,2,3}|\bar{B}_q^0\rangle=\langle B_q^0|\mathcal{O}_{1,2,3}|\bar{B}_q^0\rangle$.
This leaves five independent matrix elements.
In the Standard Model, only matrix elements of the first three operators show up in physical observables.
		
Historically, the matrix elements have been parametrized as~\cite{Lenz:2006hd, Beneke:1996gn}
\be
	\op{i} = \mathfrak{c}_i\ M_{B_q}^2\ f_{B_q}^2\ B_{B_q}^{(i)}(\mu),  \label{eq:bagparams}
\ee
with coefficients $\mathfrak{c}_i=(\nicefrac{2}{3},\ \nicefrac{-5}{12},\ \nicefrac{1}{12},\ \nicefrac{1}{2},\ \nicefrac{1}{6} )$.   
The $B_{B_q}^{(i)}$ are known as bag parameters 
and characterize the deviation from the so-called vacuum saturation-approximation~\cite{Lee:1973}.  	
Lattice QCD can now provide a direct calculation of the \me{i}.
Because of their historical use, though, most lattice calculations report the bag parameters as well as the matrix elements.

	
We next review the different matrix elements and combinations of matrix elements that appear in specific comparisons with experiment.
We look at the mass difference $\Delta M_q$, the ratio $\Delta M_s/ \Delta M_d$, and the width difference $\Delta \Gamma_q$.
The Standard Model  expression for the $B$-meson mass difference $\Delta M_q$ is given by 
\be
	\Delta M_q = 
		\left( \frac{G_F^2 M_W^2 S_0}{4 \pi^2 M_{B_q}} \right)  \eta_B(\mu)  
		|V_{tb} V_{tq}^*|^2  \op{1}.  \label{eq:massdiff} 
\ee
The quantities in parentheses and $\eta_B$ are known factors, $V_{ij}$ is a CKM matrix element, 
and the states have a relativistic normalization.
Measurements of $\Delta M_q$ have sub-percent errors~\cite{Abulencia:2006ze, Nakamura:2010zzi}.
Therefore, assuming no new physics in $B$-mixing,
constraining the CKM matrix contribution $|V_{tb} V_{tq}^*|$ relies on our ability to determine \me{1} precisely.

The ratio 
\be
	\frac{\Delta M_s}{\Delta M_d} = 
	\left| \frac{V_{ts}}{V_{td}} \right|^2
	\frac{\langle \bar{B}_s^0 | \mathcal{O}_1(\mu) | B_s^0 \rangle}{\langle \bar{B}_d^0 | \mathcal{O}_1(\mu) | B_d^0 \rangle}
	\equiv  \left| \frac{V_{ts}}{V_{td}} \right|^2 
	\frac{M_{B_s}}{M_{B_d}}\xi^2
\ee
defines $\xi$.
Some errors cancel in this ratio, making $\xi$ a more precisely determined quantity than the separate matrix elements.
Additionally, in CKM matrix fits, the use of $\xi$ can aid in minimizing correlations between lattice inputs~\cite{Laiho:2009eu}.

The $B$-meson width difference can be written as~\cite{Lenz:2006hd, Beneke:1996gn}
\be
	\Delta \Gamma_q =  
	\Big[ 
  		G_1 \;\;  \langle \bar{B}_q^0 | \mathcal{O}_1(\mu) | B_q^0 \rangle
	+ G_3 \;\; \langle \bar{B}_q^0 | \mathcal{O}_3(\mu) | B_q^0 \rangle  \;\;  \Big] \cos \phi_q
	+ O(1/m_b, \alpha_s) ,  \label{eq:deltagamma}
\ee
where the $G_i$'s are comprised of known constants and short-distance coefficients, and $\phi$ is the CP-violating phase~\cite{Lenz:2006hd}.
A calculation of the complete set of matrix elements allows one to have
results not only for the two matrix elements appearing explicitly in Eq.~(\ref{eq:deltagamma}), 
but also combinations of \me{1,2,3} that are useful to phenomenologists.
%
Specifically, $\Delta \Gamma / \Delta M$ depends on $\me{3}/\me{1}$ and the combination
$\mathcal{O}_R \equiv \mathcal{O}_2 + \mathcal{O}_3 + (1/2) \mathcal{O}_1$
is useful for estimating $1/m_b$ errors.

Finally, including Beyond-the-Standard-Model (BSM) contributions, $\Delta M_q$ takes the generic form
\be
	\Delta M_q = 	\sum_{i = 1}^5  C_i(\mu) \;  \langle B_q^0 | \mathcal{O}_i(\mu) | \overline{B}_q^0 \rangle .
\ee
Lattice values of the full set of matrix element, \me{1} through \me{5}, are needed to check that a given BSM model is consistent with experiment.
For examples, see Refs.~\cite{BSM}.

\section{The Status of Lattice Calculations}

\begin{figure}
	\begin{center}
		\begin{tabular}{cc}
		\includegraphics[width=0.4 \textwidth]{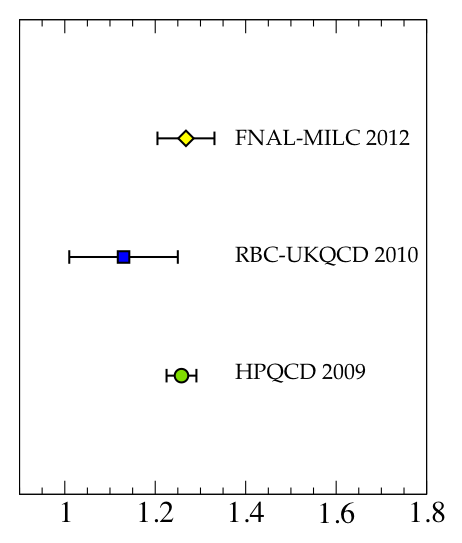}  & \includegraphics[width=0.6 \textwidth]{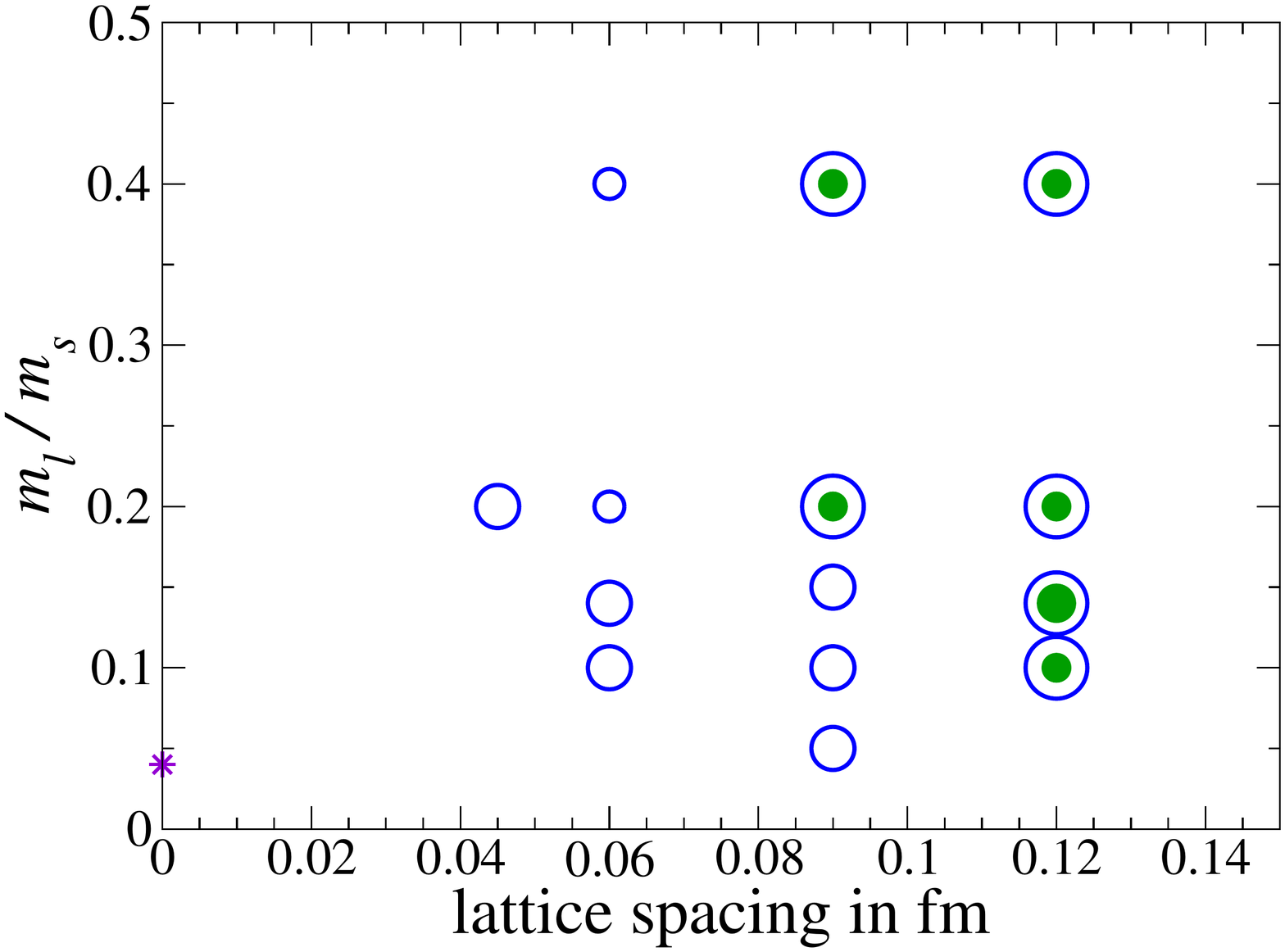} \\
		(a) & (b) \\
		\end{tabular}
		\caption{ (a) $\xi$ from three recent lattice calculations that include 2+1 sea quarks.
		The calculations are from FNAL-MILC~\cite{Bazavov:2012zs}, RBC-UKQCD~\cite{Albertus:2010nm}, and HPQCD~\cite{Gamiz:2009ku}.
			(b) A comparison of ensembles used in the calculation of Ref.~\cite{Bazavov:2012zs} 
			and the one described in these proceedings.
			The $x$ and $y$ axes show the value of the lattice spacing $a$ and the sea-quark mass ratio $m_{\rm u/d} / m_{\rm s}$, respectively.
			Filled (green) circles mark the ensembles used in Ref.~\cite{Bazavov:2012zs}. 
			Open circles mark the ensembles used in the new analysis.
			The area of each circle is approximately proportional to the amount of data on that ensemble.
			The (purple) burst at the lower left marks the physical point.}
		\label{fig:xi-compare}
	
	\end{center}
\end{figure}

The Fermilab Lattice and MILC Collaborations published the value of $\xi = 1.268(63)$ in early 2012~\cite{Bazavov:2012zs}.
This joins two other published, unquenched calculations, that of the HPQCD Collaboration~\cite{Gamiz:2009ku} 
and exploratory work done by the RBC and UKQCD Collaboration~\cite{Albertus:2010nm}.
Fig.~\ref{fig:xi-compare}~(a) compares $\xi$ from these calculations.
In the Fermilab-MILC calculation, there are two dominant sources of error.
The first is the combined error from statistics, light-quark-discretization, and chiral-continuum-extrapolation; this yields a 3.7\% error on $\xi$.
The second is an estimate of the effect of wrong-spin operators that appear in the chiral extrapolation, but were not explicitly included in the analysis; this yields a 3.2\% error on $\xi$.
The existence of wrong-spin terms was unknown at the time Ref.~\cite{Gamiz:2009ku} was published, and their effect was not addressed in that paper.
As explained in Sec.~\ref{sec:calculation} below, the error from ``wrong-spin contributions'' will not be present in our subsequent analyses.
%

We are performing the first unquenched calculation of the full set of matrix elements.
A quenched calculation by Be\'cirevi\'c et al.~\cite{Becirevic:2001xt} uses static heavy valence quarks with an interpolation to $m_b$, and a linear extrapolation in the light valence mass to reach $B_d$.
Although such approximations are no longer necessary in modern lattice-QCD calculations, only limited examples of improved calculations exist.
For $B_s$ mixing only, a one-lattice spacing, unquenched calculation of \me{1,2,3} has been done by the HPQCD Collaboration~\cite{HPQCD:O123}.
%
Several groups are actively working on calculations of the full set of matrix elements.
Our collaboration gave preliminary results in the Lattice 2011 proceedings~\cite{Bouchard:2011xj}, and the ETM Collaboration reported on preliminary work at Lattice 2012~\cite{ETMC-Lat12}.

\section{The Calculation}	\label{sec:calculation}

We first review the differences between our published calculation of $\xi$ and our current analysis of the full set of matrix elements.
We then discuss some details of the current analysis.

\subsection{Published versus current analysis}
Our calculation of $\xi$ in Ref~\cite{Bazavov:2012zs} and the work described here and previously~\cite{Bouchard:2011xj} have many points in common.
Both use the MILC gauge configurations~\cite{Bazavov:2010} with \mbox{$2+1$} flavors of asqtad staggered~\cite{stag_fermion} sea quarks.
To simulate the bottom valence quark, we use the Fermilab interpretation of the Sheikholeslami-Wohlert (clover) action~\cite{ElKhadra:1997},
with the hopping parameter $\kappa_b$ tuned to produce the observed $B_s$ meson mass~\cite{Bernard:2011}.
Light valence quarks are simulated with the asqtad action, and we work in the meson rest frame. 

Our on-going calculation of the full set of matrix elements improves on Ref.~\cite{Bazavov:2012zs} in several ways.
It includes two smaller lattice spacings, $a \approx 0.06$~and~0.045~fm, for a total of four lattice spacings.
Ensembles have sea-quark masses, with  ratios of $m_{\rm u/d} / m_{\rm s} = 0.1$ at all lattice spacings 
and one ensemble with $m_{\rm u/d} / m_{\rm s} = 0.05$.
On that ensemble, the light valence-quark mass range has been extended to include $0.05 m_{\rm s}$.
Additionally, statistics on ensembles used in the earlier calculation have been increased by a factor of about four.
This information is summarized in Fig.~\ref{fig:xi-compare}~(b).

Our current calculation uses complete expressions for the chiral perturbation theory as described below and in Ref.~\cite{CBernard-chipt}.
This includes contributions from wrong-spin terms, and removes the need for the ``wrong-spin contributions'' error that appears for $\xi$ in Ref.~\cite{Bazavov:2012zs}.
%

In addition to the five matrix elements, 
our current calculation includes an evaluation of the bag parameters for each operator.
Because our collaboration is calculating the decay constants $f_{B_{q}}$~\cite{Neil-Lat12}, we will be able to account for correlations between the two results when extracting the these parameters.
Finally, combinations of the matrix elements including $\xi$, the ratio $\me{3}/\me{1}$, and $\mathcal{O}_R$ will be determined.

\subsection{Details of the new analysis}

We generate two-point meson correlation functions as well as three-point correlation functions for the operators in Eq.~(\ref{eq:susybasis}).
For each meson, the two- and three-point data is fit simultaneously, with a common value for the $B$-meson energy, using constrained curve fitting~\cite{Lepage:2001ym, Wingate:2003}.
Simultaneous fits easily account for correlations between the two- and three-point data, require fewer fits, and allow the two-point data to help constrain the energies, resulting in more precise fit results for the three-point amplitudes.
Explicit operator expressions and fit functions can be found in Ref.~\cite{Bouchard:2011xj}.

With the three-point results in hand, we renormalize the matrix elements.
These mix under renormalization, e.g.
\be
	\langle \mathcal{O}_1 \rangle^R = 
	(1 + \alpha_s \zeta_{11}) \langle \mathcal{O}_1 \rangle + \alpha_s \zeta_{12}	\langle \mathcal{O}_2 \rangle.
\ee
The $\zeta_{ij}$ are calculated using one-loop, mean-field improved lattice perturbation theory~\cite{Gamiz:inprep}. 
For $\alpha_s$, we use the ``V'' scheme~\cite{Lepage:1993} with four-loop running, as implemented in Ref.~\cite{Mason:2005}, and take $\alpha_s = \alpha_{\rm V}(2/a)$.

The last step in the analysis is the chiral-continuum extrapolation.
We use SU(3), partially-quenched, heavy-meson, staggered \cpt~\cite{CBernard-chipt, Aubin:2005aq}, which generalizes the continuum calculation of Ref.~\cite{Detmold:2007} to include staggered discretization errors.
With staggered light quarks, matrix elements of wrong-spin operators appear in the \cpt~\cite{CBernard-chipt}.  
These contributions vanish in the continuum, but can be of comparable size to other NLO chiral effects at non-zero lattice spacing.
Because the five matrix elements $\langle \mathcal{O}_{1\dots 5} \rangle$ form a complete basis, 
wrong-spin contributions can be written in terms of them.
This, fortunately, means no new low-energy constants (LECs) are introduced.
Mixing occurs among  
$\langle \mathcal{O}_1 \rangle$, $\langle \mathcal{O}_2 \rangle$, $\langle \mathcal{O}_3 \rangle$, 
and, separately, among $\langle \mathcal{O}_4 \rangle$, $\langle \mathcal{O}_5 \rangle$.
As an example of the mixing, the expression for \me{1} at next-to-leading order is
\bea
	\langle \overline{B}_q^0|{O}_1^q|B_q^0 \rangle =&&
	\beta_1\Bigg( 1+
	\frac{{W}_{q\overline{b}}+{W}_{b\overline{q}}}{2} + T_q  + Q_q
	+ \tilde{T}^{(\rm{a})}_q  + \tilde{Q}^{({\rm a})}_q    \Bigg)  \nonumber \\
	&& + (2\beta_2+2\beta_3) \tilde{T}^{(\rm{b})}_q 
	 + (2\beta'_2+2\beta'_3) \tilde{Q}^{({\rm b})}_q 
	 + \text{NLO analytic terms}.	\label{eq:chipt}
\eea
The terms $W$, $T$, and $Q$, are ``correct spin'' contributions from wave-function renormalization, tadpole diagrams, and sunset diagrams, respectively.
Terms $\tilde{T}$ and $\tilde{Q}$, from tadpole and sunset diagrams, are contributions from wrong-spin operators.
The $\beta_i$ and $\beta'_i$ are the leading-order LEC's for the matrix element \me{i}.
Equation~(\ref{eq:chipt}) shows explicitly that no new LEC's appear in the chiral expression.
In our analysis, we perform simultaneous fits to each group of matrix elements that mix.

\section{Outlook}

We have recently completed a calculation of the SU(3)-breaking ratio $\xi$ used in the study of neutral $B$ mixing~\cite{Bazavov:2012zs}.
In these proceedings, we present an update of our $2+1$-flavor calculation of the five matrix elements needed to describe neutral $B$ mixing in and beyond the Standard Model.
This expanded calculation improves on the methods used in our calculation of $\xi$ via extended data sets, increased statistics, and improved analysis, particularly in the chiral-continuum extrapolation.  
Preliminary results from this analysis can be found in Ref.~\cite{Bouchard:2011xj}.

\section*{Acknowledgments}
%
Computations for this work were carried out with resources provided by
the USQCD Collaboration, the Argonne Leadership Computing Facility,
the National Energy Research Scientific Computing Center, and the 
Los Alamos National Laboratory, which are funded by the Office of Science of the
U.S. Department of Energy; 
and with resources provided by the National Institute for Computational Science, 
the Pittsburgh Supercomputer Center, the San Diego Supercomputer Center, 
and the Texas Advanced Computing Center, 
which are funded through the National Science Foundation's Teragrid/XSEDE Program.
This work was supported in part by the U.S. Department of Energy under 
Grants No.~DE-FG02-91ER40677 (C.M.B, E.D.F., A.X.K.) and No.~DE-FG02-91ER40628 (C.B.),
and by the Fermilab Fellowship in Theoretical Physics (C.M.B.).
This work was supported in part by the MICINN (Spain) under grant FPA2010-16696 and \emph{Ram\'on y Cajal} program (E.G.), Junta de Andaluc\'{\i}a (Spain) under grants FQM-101, FQM-330, and FQM-6552 (E.G.), European Commission under Grant No. PCIG10-GA-2011-303781 (E.G.)
 Fermilab is operated by Fermi Research Alliance, LLC, under Contract
No.~DE-AC02-07CH11359 with the United States Department of Energy.



\begin{thebibliography}{99}

 
\bibitem{Bazavov:2012zs} 
  A.~Bazavov {\em et al.} [Fermilab Lattice and MILC], 
  Phys.\ Rev.\ D {\bf 86}, 034503 (2012)
  [arXiv:1205.7013 [hep-lat]].
  
  
\bibitem{Laiho:2009eu}  
  J.~Laiho, E.~Lunghi and R.S.~Van~de~Water,
  Phys.\ Rev.\  D {\bf 81}, 034503 (2010)
  [\href{http://arxiv.org/abs/0910.2928}{arXiv:0910.2928}].

\bibitem{Lunghi:2009ke} 
  E.~Lunghi and A.~Soni,
  Phys.\ Rev.\ Lett.\  {\bf 104}, 251802 (2010)
  [\href{http://arxiv.org/abs/0912.0002}{arXiv:0912.0002}].

\bibitem{Laiho:2011nz} 
  J.~Laiho, E.~Lunghi and R.S.~Van~de~Water,
  PoS {\bf FPCP2010}, 040 (2010)
  [\href{http://arxiv.org/abs/1102.3917}{arXiv:1102.3917}].
 
\bibitem{Lenz:2010gu}  
  A.~Lenz {\it et al.},
  Phys.\ Rev.\  D {\bf 83}, 036004 (2011)
  [\href{http://arxiv.org/abs/1008.1593}{arXiv:1008.1593}].
  
\bibitem{Bona:2008jn}
  M.~Bona {\it et al.}  [UTfit],
  PMC Phys.\  A {\bf 3}, 6 (2009)
  [\href{http://arxiv.org/abs/0803.0659}{arXiv:0803.0659}].
  

\bibitem{Lenz:2012az} 
  A.~Lenz {\em et al.}, 
  Phys.\ Rev.\ D {\bf 86}, 033008 (2012)
  [arXiv:1203.0238 [hep-ph]].
  
  
\bibitem{CKM-Nierste}
	U.~Nierste,  
	``$B$ mixing in the Standard Model and Beyond'',
	7th International Workshop on the CKM Unitarity Triangle, Cincinnati, Ohio, USA.
  
\bibitem{CKM-Summary-HM}
 	S.~Hansmann-Menzemer, U.~Nierste, F.~Wilson;  
	``WG IV Summary: Mixing and mixing-related CP violation in $B$ system'',
	7th International Workshop on the CKM Unitarity Triangle, Cincinnati, Ohio, USA.
   
\bibitem{Lenz:2006hd}
  A.~Lenz and U.~Nierste,
  JHEP {\bf 0706}, 072 (2007)
  [\href{http://arxiv.org/abs/hep-ph/0612167}{arXiv:hep-ph/0612167}].


\bibitem{Beneke:1996gn}
  M.~Beneke, G.~Buchalla and I.~Dunietz,
  Phys.\ Rev.\  D {\bf 54}, 4419 (1996)
  [Erratum-ibid.\  D {\bf 83}, 119902 (2011)]
  [\href{http://arxiv.org/abs/hep-ph/9605259}{arXiv:hep-ph/9605259}].
  
  
 
  
\bibitem{Lee:1973}
  B.W.~Lee, J.R.~Primack, and S.B.~Treiman,
  Phys.\ Rev.\ D {\bf 7}, 510 (1973).


  

  \bibitem{Abulencia:2006ze}
  A.~Abulencia {\it et al.} [CDF],
  Phys.\ Rev.\ Lett.\  {\bf 97}, 242003 (2006)
  [\href{http://arxiv.org/abs/hep-ex/0609040}{arXiv:hep-ex/0609040}].

\bibitem{Nakamura:2010zzi}
  K.~Nakamura {\it et al.} [Particle Data Group],
  J.\ Phys.\ G {\bf 37}, 075021 (2010) and 2011 partial update for the 2012 edition
  [\href{http://pdg.lbl.gov}{http://pdg.lbl.gov}].
 
 
 
  
  
  

\bibitem{BSM}
  W.~Altmannshofer and M.~Carena,
  [\href{http://arxiv.org/abs/1110.0843}{arXiv:1110.0843}];
  M.~Bona {\it et al.}  \mbox{[UTfit]},
  JHEP {\bf 0803}, 049 (2008)
  [\href{http://arxiv.org/abs/0707.0636}{arXiv:0707.0636}];
  B.A.~Dobrescu and G.Z.~Krnjaic,
  [\href{http://arxiv.org/abs/1104.2893v1}{arXiv:1104.2893}].


  
  
  
\bibitem{Gamiz:2009ku}
  E.~G\'amiz {\it et al.} [HPQCD],
  Phys.\ Rev.\  D {\bf 80}, 014503 (2009)
  [\href{http://arxiv.org/abs/0902.1815}{arXiv:0902.1815}].
 
\bibitem{Albertus:2010nm} 
  C.~Albertus {\em et al.} [RBC and UKQCD], 
  Phys.\ Rev.\ D {\bf 82}, 014505 (2010)
  [arXiv:1001.2023 [hep-lat]].
  

	

\bibitem{Becirevic:2001xt}
  D.~Be\'cirevi\'c {\it et al.},
  JHEP {\bf 0204}, 025 (2002)
  [\href{http://arxiv.org/abs/hep-lat/0110091}{arXiv:hep-lat/0110091}].



\bibitem{HPQCD:O123}
  
	J.~Shigemitsu {\it et al.} [HPQCD],
	\href{http://pos.sissa.it/archive/conferences/032/093/LAT2006_093.pdf}{PoS {\bf Lattice 2006} 093}.


  
\bibitem{Bouchard:2011xj} 
  C.~M.~Bouchard {\em et al.} [Fermilab Lattice and MILC], 
  PoS LATTICE {\bf 2011}, 274 (2011)
  [arXiv:1112.5642 [hep-lat]].


\bibitem{ETMC-Lat12}
  N.~Vela [European Twisted Mass], PoS LAT2012, 105 (2012).



%
%
%
%


\bibitem{Bazavov:2010}
	A.~Bazavov {\it et al.} [MILC],
	Rev. Mod. Phys. {\bf 82}, 1349 (2010)
	[\href{http://arxiv.org/abs/0903.3598}{arXiv:0903.3598}].





  \bibitem{stag_fermion}
  T.~Blum {\it et al.},       
  Phys.\ Rev.\ D {\bf 55}, R1133 (1997)
  [arXiv:hep-lat/9609036];
%
  K.~Orginos and D.~Toussaint  [MILC Collaboration],
  Phys.\ Rev.\ D {\bf 59}, 014501 (1998)
  [arXiv:hep-lat/9805009];
%
  J.~F.~Laga\"e and D.~K.~Sinclair,
  Phys.\ Rev.\ D {\bf 59}, 014511 (1998)
  [arXiv:hep-lat/9806014];
%
  G.~P.~Lepage,        
  Phys.\ Rev.\ D {\bf 59}, 074502 (1999)
  [arXiv:hep-lat/9809157];
%
  K.~Orginos, D.~Toussaint and R.~L.~Sugar  [MILC Collaboration],      
  Phys.\ Rev.\ D {\bf 60}, 054503 (1999)
  [arXiv:hep-lat/9903032];
%
  C.~W.~Bernard {\it et al.}  [MILC Collaboration],
  Phys.\ Rev.\ D {\bf 61}, 111502(R) (2000)
  [arXiv:hep-lat/9912018].




\bibitem{ElKhadra:1997}
	A.X.~El-Khadra, A.S.~Kronfeld, and P.B.~Mackenzie,
	Phys. Rev. D {\bf 55}, 3933 (1997)
	[\href{http://arxiv.org/abs/hep-lat/9604004}{arXiv:hep-lat/9604004}].

\bibitem{Bernard:2011}
	C.~Bernard {\it et al.} [Fermilab Lattice and MILC],
	Phys. Rev. D {\bf 83}, 034503 (2011)
	[\href{http://arxiv.org/abs/1003.1937}{arXiv:1003.1937}].
	
	

	
\bibitem{CBernard-chipt}
	C.~Bernard [MILC], PoS LAT2012, 201 (2012), and in preparation.
	
	
\bibitem{Neil-Lat12}
	E.~T.~Neil, C.~Bernard, J.~N.~Simone [Fermilab Lattice and MILC], PoS LAT2012, 101 (2012). 


\bibitem{Lepage:2001ym} 
  G.P.~Lepage {\it et al.},
  Nucl.\ Phys.\ Proc.\ Suppl.\  {\bf 106}, 12 (2002)
  [\href{http://arxiv.org/abs/hep-lat/0110175}{arXiv:hep-lat/0110175}].
  
\bibitem{Wingate:2003}
	M.~Wingate {\it et al.},
	Phys. Rev. D {\bf 67}, 054505 (2003)
	[\href{http://arxiv.org/abs/hep-lat/0211014}{arXiv:hep-lat/0211014}].




	






%



\bibitem{Gamiz:inprep}
	E.~G\'amiz {\it et al.} [HPQCD],
  	Phys.\ Rev.\ D {\bf 77}, 114505 (2008)
  	[\href{http://arxiv.org/abs/0804.1557}{arXiv:0804.1557}].;
	E.~G\'{a}miz, A.X.~El-Khadra, and A.S.~Kronfeld,
	{\it in preparation.}


\bibitem{Lepage:1993}
	G.P.~Lepage and P.B.~Mackenzie,
	Phys. Rev. D {\bf 48}, 2250 (1993)
	[\href{http://arxiv.org/abs/hep-lat/9209022}{arXiv:hep-lat/9209022}].


\bibitem{Mason:2005}    
	Q.~Mason {\it et al.},
	Phys. Rev. Lett. {\bf 95}, 052002 (2005)
	[\href{http://arxiv.org/abs/hep-lat/0503005}{arXiv:hep-lat/0503005}].
	
	
	
%
%
  

	
\bibitem{Aubin:2005aq} 
  C.~Aubin and C.~Bernard,
  Phys.\ Rev.\ D {\bf 73}, 014515 (2006)
  [hep-lat/0510088]

\bibitem{Detmold:2007}
	W.~Detmold and C.-J.~David~Lin,
	Phys. Rev. D {\bf 76}, 014501 (2007)
	[\href{http://arxiv.org/abs/hep-lat/0612028}{arXiv:hep-lat/0612028}].
	
	

%



	

  

\end{thebibliography}
\end{document}